# A Radiomics-Incorporated Deep Ensemble Learning Model for Multi-Parametric MRI-based Glioma Segmentation

Yang Chen[1], Zhenyu Yang[2], Jingtong Zhao[2], Justus Adamson[2], Yang Sheng[2], Fang-Fang Yin[2], Chunhao Wang[2*]

[1]Medical Physics Graduate Program, Duke Kunshan University, Kunshan, Jiangsu, China 215316
[2]Department of Radiation Oncology, Duke University, Durham, NC, 27710

**Short Running Title: Radiomics Deep Ensemble for Glioma Segmentation**

*Corresponding authors:
Chunhao Wang, Ph.D.
Box 3295, Department of Radiation Oncology
Duke University Medical Center
Durham, NC, 27710, United States
E-mail: chunhao.wang@duke.edu

Funding Statement: None associated with this work.

Conflict of Interest: None associated with this work.

Abstract

*Purpose:*

To develop a deep ensemble learning model with a radiomics spatial encoding execution for improved glioma segmentation accuracy using multi-parametric MRI (mp-MRI).

*Methods:*

This model was developed using 369 glioma patients with a 4-modality mp-MRI protocol: T1, contrast-enhanced T1 (T1-Ce), T2, and FLAIR. In each modality volume, a 3D sliding kernel was implemented across the brain to capture image heterogeneity: fifty-six radiomic features were extracted within the kernel, resulting in a $4^{th}$ order tensor. Each radiomic feature can then be encoded as a 3D image volume, namely a radiomic feature map (RFM). For each patient, all RFMs extracted from all 4 modalities were processed by the Principal Component Analysis (PCA) for dimension reduction, and the first 4 principal components (PCs) were selected. Next, four deep neural networks as sub-models following the U-Net's architecture were trained for the segmenting of a region-of-interest (ROI): each sub-model utilizes the mp-MRI and 1 of the 4 PCs as a 5-channel input for 2D execution. Last, the 4 softmax probability results given by the U-net ensemble were superimposed and binarized by Otsu's method as the segmentation result. Three ensemble models were trained to segment enhancing tumor (ET), tumor core (TC), and whole tumor (WT), respectively. Segmentation results given by the proposed ensemble were compared to the mp-MRI-only U-net results.

*Results:*

All 3 radiomics-incorporated deep ensemble learning models were successfully implemented: comparing to mp-MRI-only U-net results, the dice coefficients of ET (0.777→0.817), TC (0.742 → 0.757), and WT (0.823 → 0.854) demonstrated improvements. Accuracy, sensitivity, and specificity results demonstrated the same patterns.

*Conclusion:*

The adopted radiomics spatial encoding execution enriches the image heterogeneity information that leads to the successful demonstration of the proposed deep ensemble model, which offers a new tool for mp-MRI based medical image segmentation.

# 1. Introduction

Gliomas are the most common primary tumors in the brain and spinal cord, with a poor prognosis (Claus *et al.*, 2015; Chen *et al.*, 2017; Hanif *et al.*, 2017). An accurate delineation of the glioma and its sub-volumes is essential to improve the treatment results and prognosis (ABDALLA; *et al.*, 2020; Sethi *et al.*, 2011). For many years, magnetic resonance imaging (MRI) has been considered the gold standard of radiographic glioma diagnosis since the MRI can provide highly specific imaging of the soft tissues (ABDALLA; *et al.*, 2020; Sethi *et al.*, 2011; Zhang *et al.*, 2021). While single modality imaging has been reported with limited specificity (Demirel and Davis, 2018), multi-parametric MRI (mp-MRI) with T1 weighted (T1), T2 weighted (T2), T2 Fluid-Attenuated Inversion Recovery (FLAIR), and T1 contrast-enhanced (T1-Ce) sequences can give more accurate assessments of the glioma tissue microenvironment (ABDALLA; *et al.*, 2020; Demirel and Davis, 2018; Upadhyay and Waldman, 2011).

Currently, the delineation of gliomas is predominantly performed by hand, which is a tedious and time-intensive process (Zhang *et al.*, 2021; Ghaffari *et al.*, 2020). The results of manual delineation, however, can also vary significantly among different human raters due to different experiences and preferences (Deeley *et al.*, 2011; Deeley *et al.*, 2013; Mazzara *et al.*, 2004). Many recent studies have thus been done to improve the accuracy and efficiency of glioma segmentation (Işın *et al.*, 2016). Deep learning, as a newly emerging computational method, has become a powerful tool within the intersection field of computer science and medical imaging (Di Ieva *et al.*, 2021). Pilot studies have demonstrated deep learning's potential in promising glioma segmentation accuracy with high efficiency in the pre-clinical setting (Di Ieva *et al.*, 2021; Lachinov *et al.*, 2019; Perkuhn *et al.*, 2018); Nevertheless, more works need to be done to further improve the segmentation accuracy towards potential clinical application (Isensee *et al.*, 2021; Lotan *et al.*, 2019). While the majority of the recent and current relevant works focused on the improvements of neural network architecture designs, research works about neural network input optimization are relatively limited (Akkus *et al.*, 2017; Hesamian *et al.*, 2019; Magadza and Viriri, 2021; Tiwari *et al.*, 2020).

Radiomics is a recent advance of computational imaging analysis that extracts high-dimensional mineable information from clinical imaging, and it has become a major computational approach for image-based disease diagnosis, treatment effect analysis, and outcome prediction (Kumar *et al.*, 2012; Rizzo *et al.*, 2018; Lambin *et al.*, 2012; Gillies *et al.*, 2016). As handcraft computational biomarkers, radiomic features are designed to capture image intensity heterogeneity in a pre-defined spatial region-of-interest (ROI) (Gillies *et al.*, 2016; Holbrook *et al.*, 2020; Blumenthal *et al.*, 2017). Patient-specific radiomic signature is formed by concatenating individual radiomic features as a vector. As an ROI-based approach, this vector representation fails to capture voxel-wise image textural information. Our group recently invented a radiomic spatial encoding method: a 3D sliding window kernel was implemented throughout the entire image view, while radiomic features extracted within the kernel can be encoded to a voxel centered at the kernel (Yang; *et al.*, 2022b). Thus, each radiomic feature can be extended from a vector to a volumetric image with the same dimension, i.e., radiomic feature map (RFM); patient-specific radiomic signature thus becomes a $4^{th}$ tensor.

In this work, we propose a radiomics-incorporated neural network ensemble for improved glioma segmentation using mp-MRI. We hypothesize that RFMs by the radiomic spatial encoding method will enrich the mp-MRI spatial heterogeneity information that will benefit the glioma segmentation. An ensemble learning scheme was adopted, in which each sub-model, i.e., deep neural network, utilizes a unique combination of mp-MRI and RFM expression as the input: this intends to investigate the potentially complementary inclusion of different RFM expressions. Several comparison studies are included to demonstrate the merits of the proposed ensemble model.

## 2. Materials and Methods

### A. Patient Data

The Brain Tumor Segmentation Challenge 2020 (BraTS 2020) training dataset with 369 glioma patients' data was adopted by this work (Bakas; *et al.*, 2019). For each patient, the pre-resection mp-MRI includes 4 sequences, T1, T1ce, T2, and FLAIR, to fully assess the cranial anatomy for gross resection determination (Henry *et al.*, 2021). All MR images were resampled to the same spatial resolution ($1 \times 1 \times 1 mm^3$) with skull stripping after human readers' annotations (Bakas; *et al.*, 2019). The annotations include the Gd-enhancing tumor (ET), the peritumoral edema (ED), and non-enhancing tumor core (NCR/NET). This research work focused on the segmentation of three tumor regions, including enhancing tumor (ET), tumor core (TC), and whole tumor (WT): TC is the Boolean sum of ET and NCR/NET, and WT is the Boolean sum of ET, ED, and NCR/NET. Figure 1 below illustrates the ET, TC and WT segmentations of an example case.

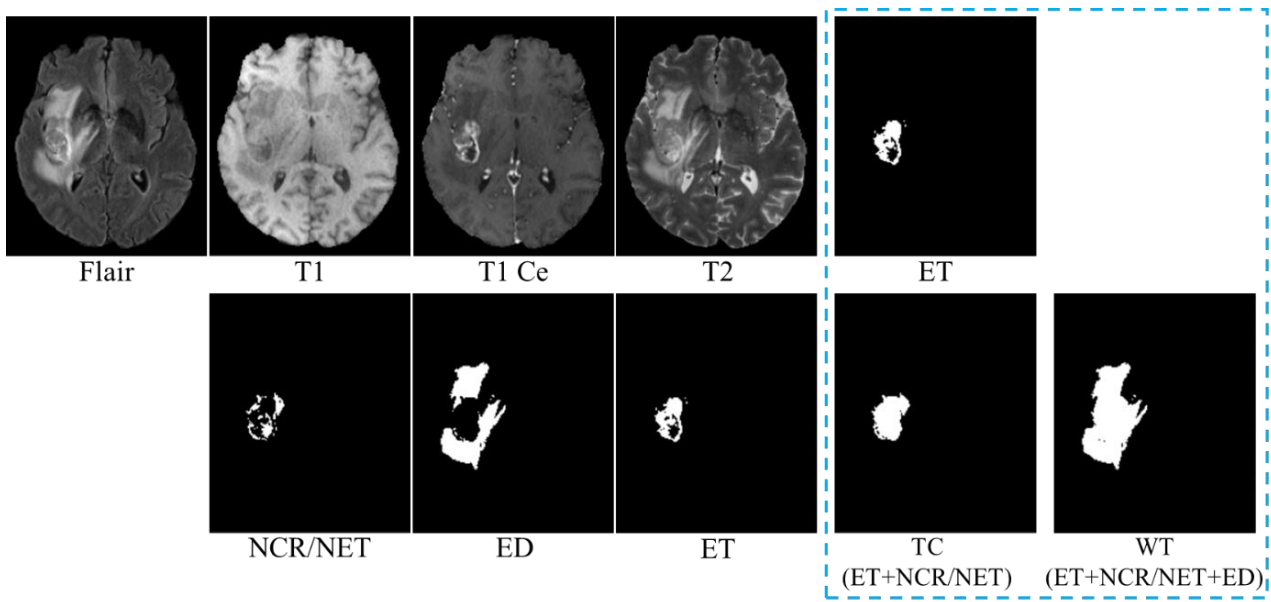

*Figure 1: An example of the mp-MRI, original segmentations (NCR/NET, ED, and ET), and the adopted ET, TC, and WT segmentations of an example patient.*

## B. Deep Ensemble Model Design

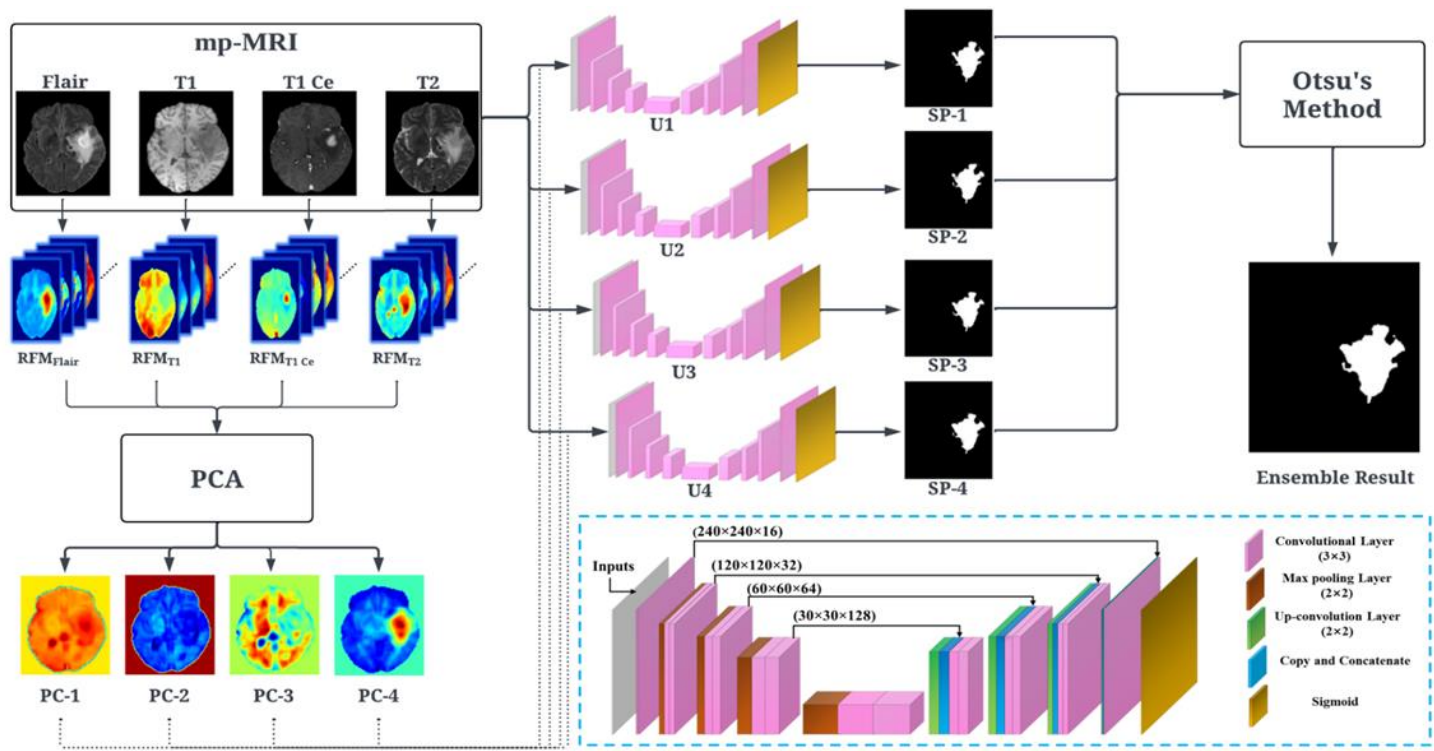

*Figure 2: The overall workflow of the proposed ensemble design. The four sub-models (U1 – U4) follow the U-net architecture specified in the blue box with detailed dimension and structure information. The 4 softmax probability maps (SP-1 – SP-4) are superimposed before thresholding by the Otsu's Method*

Figure 2 provides the overall workflow of our deep ensemble model design for segmenting a specific ROI. The RFMs in a spatial-encoded radiomics expression were calculated following our previous work (Yang; *et al.*, 2022b): for an entire 3D MR volume, a 3D sliding window kernel was implemented. For each image voxel, 56 radiomic features, including 22 features from Gray-Level Co-Occurrence Matrix (GLCOM) (H, 1973), 16 features from Gray-Level Run-Length Matrix (GLRLM) (Haralick, 1979), and 18 first-order statistics, were calculated within the region-of-interest of kernel definition. After the spatial filtering, each radiomic feature was extended as a 3D image volume (i.e., RFM) with the same original MR volume dimension. Thus, a total of 224 RFMs (56 x 4 modalities) were acquired for each patient. Given the patient cohort size (369) and number of RFMs (224) of each patient, an efficient data dimension reduction strategy is in demand to implement a practical deep learning design using RFMs without curse of dimensionality. In this work, we adopted principal components analysis (PCA): for each patient, we extracted the first four principal components (PCs) with an explained variance ratio threshold of 95%. Figure 2 provides a showcase of 4 derived RFM PCs at its lower left region.

The ensemble model includes 4 sub-models of deep neural networks (DNNs) following the 2D U-Net architecture (Yang; *et al.*, 2022a), namely U1 – U4. A detailed network structure of each sub-model with dimension information can be found at the bottom of Figure 2: an encoding path followed by a decoding path aim to extract the complex deep features from the input images and resample the imaging back to the original size (Jiang *et al.*, 2020). All the activation functions in the encoding and decoding paths were set as the Rectified Linear Unit (ReLU) activation, and the activation function of the output layer is set as the sigmoid.

Following the bagging strategy, U1 – U4 were trained independently with different input designs: each sub-model's input has 5 channels, which include 4 axial image slices from mp-MRI and one of the four RFM PCs (namely PC-1 – PC-4). Results of U1 – U4 were 4 softmax probability maps with pixel values between 0 and 1. The binarized segmentation result from each sub-model could be acquired by thresholding the softmax probability map. To generate an ensemble segmentation result that combines the outputs of all sub-models, we utilized the Otsu's method (Otsu, 1979) to threshold the superimposed softmax probability map and obtain a binarized segmentation mask

### C. Model training and comparison studies

In this work, the 369 patients was randomly split into the training (n = 295) and the test (n = 74) dataset followed by an 8:2 ratio. Three ensembles were trained for the ET, TC, and WT segmentation, respectively. Prior to the ensemble model training, all images were first normalized to [0-255] range as a uint8 format. The model training was conducted on the TensorFlow 2.5.0 platform based on the Keras library. The binary cross entropy (BCE) was adopted as the loss function (Cybenko *et al.*, 1998), and the learning rate was set at 1e-5 using the Adam optimizer. Our hardware platform was a Nvidia RTX A6000 GPU with 48 GB memory, and the data batch size was set to 80 accordingly.

The segmentation result was evaluated by the voxel-wise sensitivity, specificity, accuracy, and dice coefficients of the binarized segmentation mask using the test sets. Additionally, two other segmentation models were studied:

1) The U-Net model using mp-MRI as the only input data (*'mp-MRI-only'*).
2) The U-Net model using mp-MRI and 4 RFM PCs in a single 8-channel input design (*'mp-MRI+4 PCs'*).

In both models, the same U-net architecture (except input dimension) shown in Figure 2 was employed. The other training parameter settings remained the same. The Wilcoxon signed-rank test was adopted to examine the potential differences among the three studied models with a significant level of 0.05.

## 3. Results

Figure 3 shows the normalized mutual information (NMI) values between RFM PCs and the *mp-MRI-only* model's deep features after each convolution layer. As illustrated, increasing trends were found in the majority of the plots, suggesting that the derived RFM PCs from mp-MRI has noticeable correlations with the U-net's deep features that are also derived from mp-MRI. This observation resonates with our hypothesis that RFMs may contribute to improved image segmentation accuracy when they are incorporated into the deep learning input design.

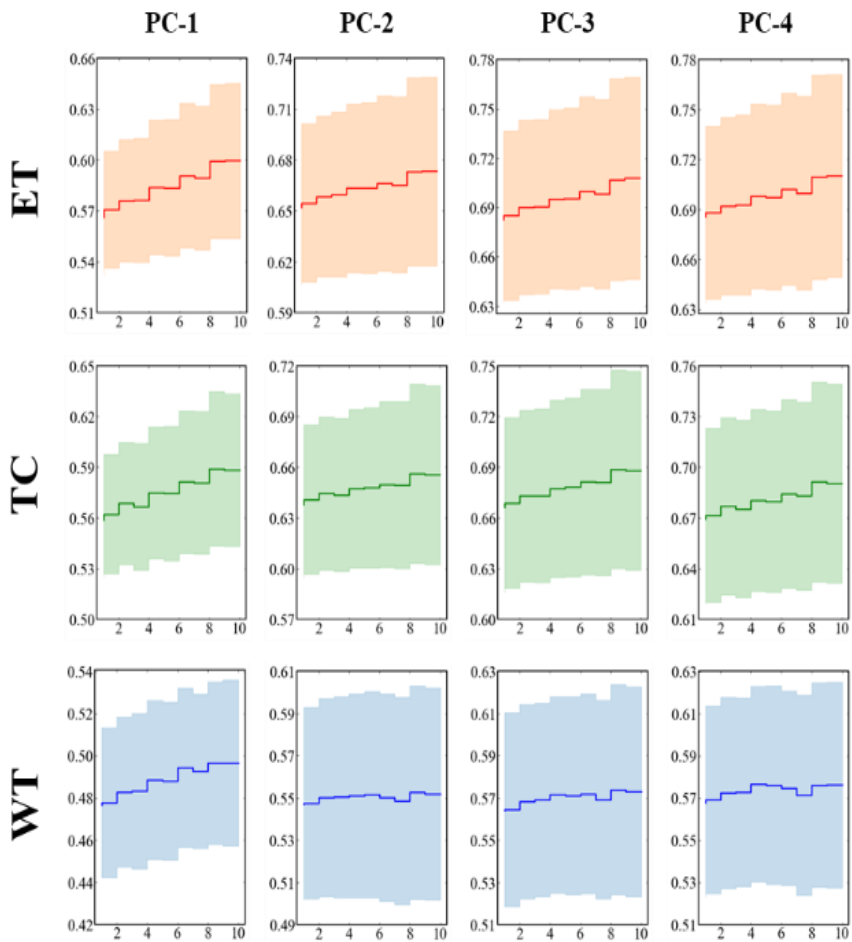

*Figure 3: Normalized mutual information between the 4 RFM PCs and vanilla U-Net deep features as a function of U-net layers (horizontal axis). The solid lines and the color shades represent mean/variance of all training set patients' results.*

Figure 4 shows an example of the whole tumor (WT) segmentation results. As shown, the four derived RFM PCs captured distinct image contrasts from the four mp-MRI images and exhibited greater intensity variations. Although the softmax probability maps of the four sub-models yielded results that were similar to the ground truth with some minor differences, the ensemble result achieved the most accurate segmentation. As an example, the ensemble result successfully captured the hook-shaped detail indicated by the red circle on the right side, which was missed by the *mp-MRI-only* model. The *mp-MRI + 4 PCs* model also achieved good segmentation accuracy but it captured some false positive signals (indicated by the orange arrow). This finding emphasizes the advantages of an ensemble learning design.

Figures 5 and 6 display the segmentation results for the TC and ET regions, respectively. Overall, the proposed deep ensemble model demonstrated the highest segmentation accuracy compared to the ground truth results.

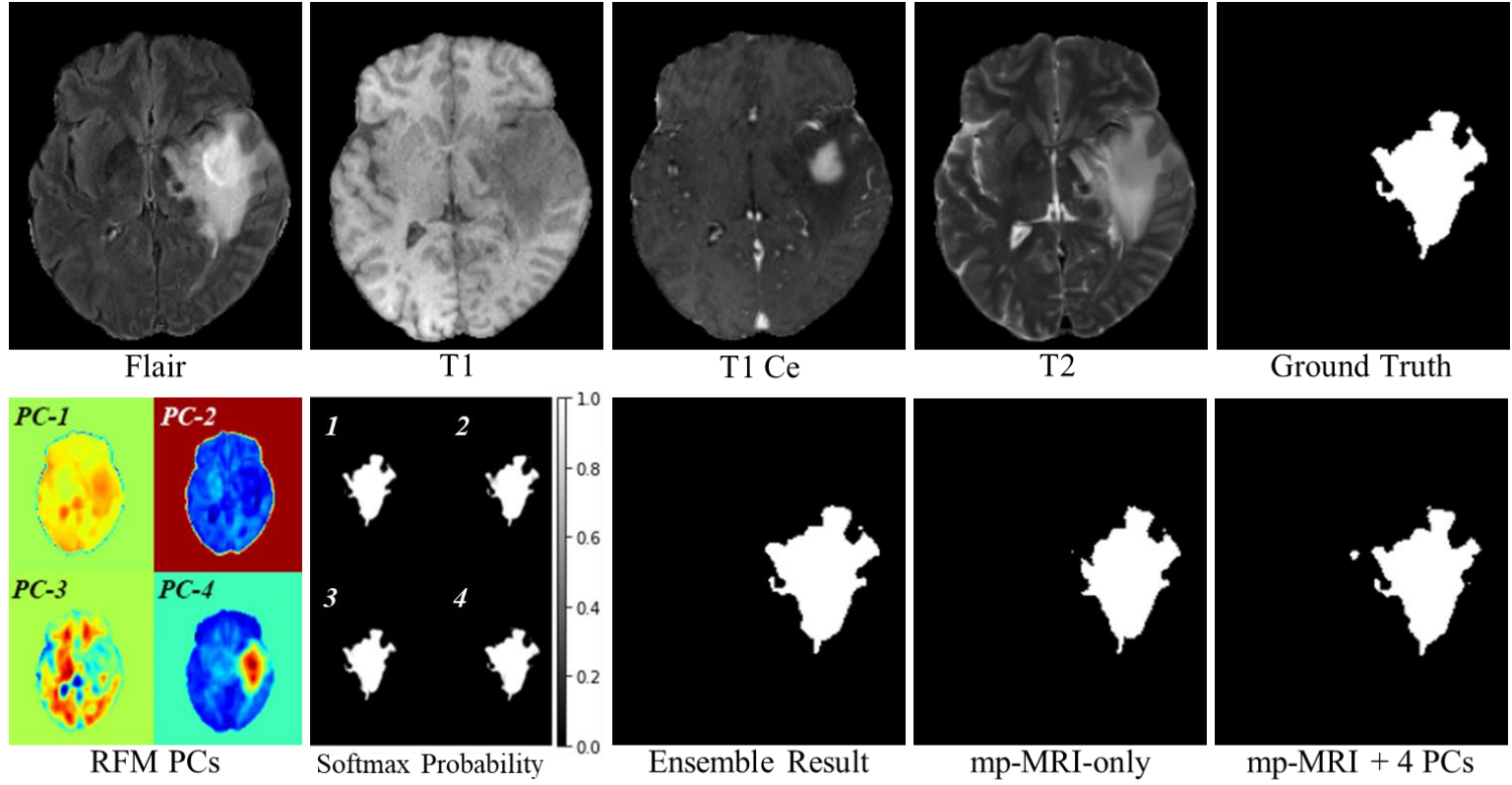

*Figure 4: An example of whole tumor (WT) segmentation comparison study results.*

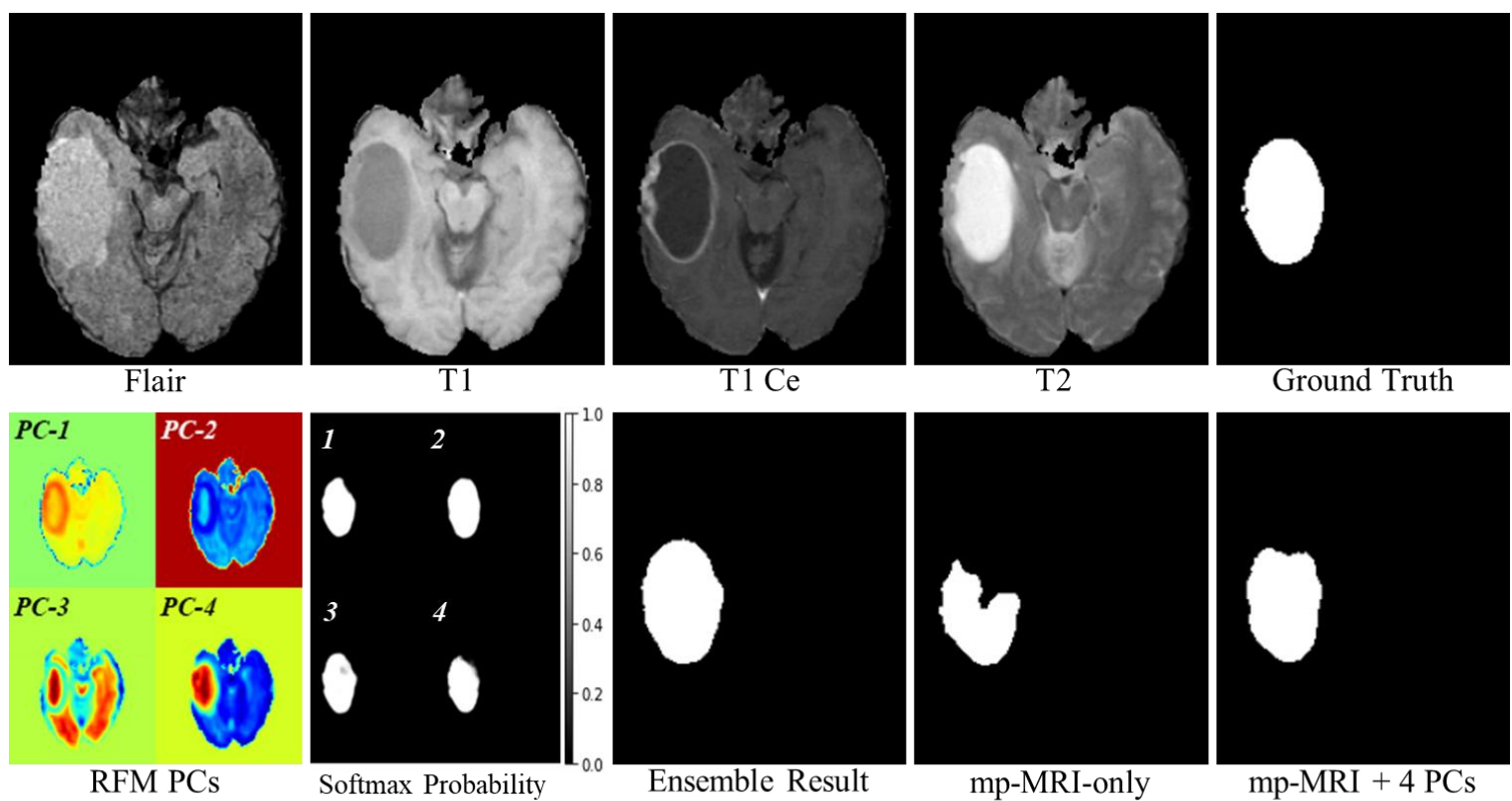

*Figure 5: An example of tumor core (TC) segmentation comparison study results.*

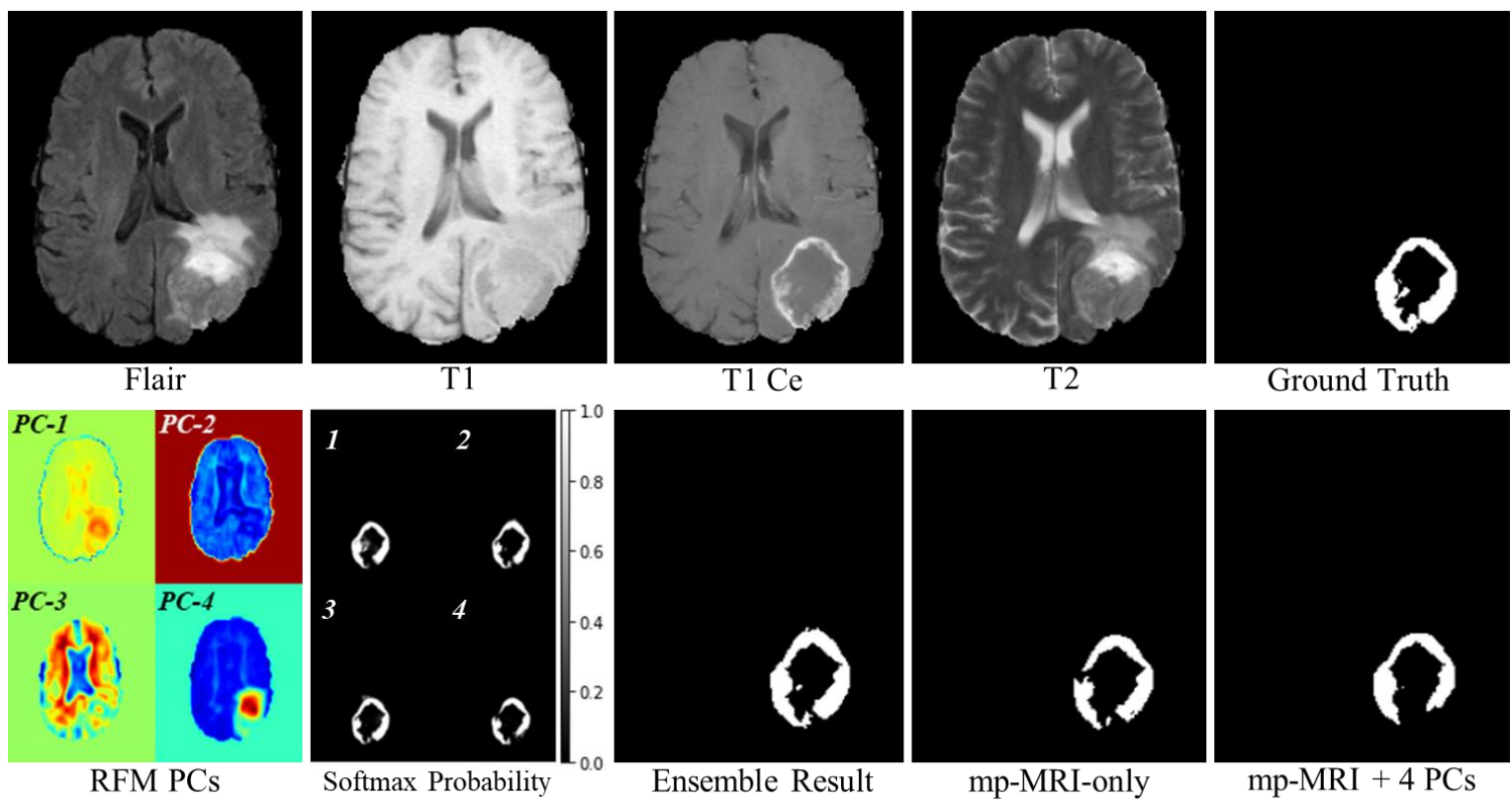

*Figure 6: An example of enhancing tumor (ET) segmentation comparison study results.*

Figure 7 shows the dice coefficients of the three models in the comparison studies above. It is evident that the deep ensemble learning model achieved the highest median dice coefficient with the lowest variance in ET and WT segmentations; its improvement in TC segmentation was found to be relatively limited. Table 1 summarizes the results of all numerical segmentation evaluators. The proposed deep ensemble model demonstrated improved dice coefficient results, along with significant increases in voxel-wise sensitivity. Additionally, the *mp-MRI + 4 PCs* model showed improved ET and WT segmentations compared to the mp-MRI-only model, while achieving comparable results in TC segmentations. These findings highlight the importance of both RFM inclusion and ensemble learning in achieving high segmentation accuracy.

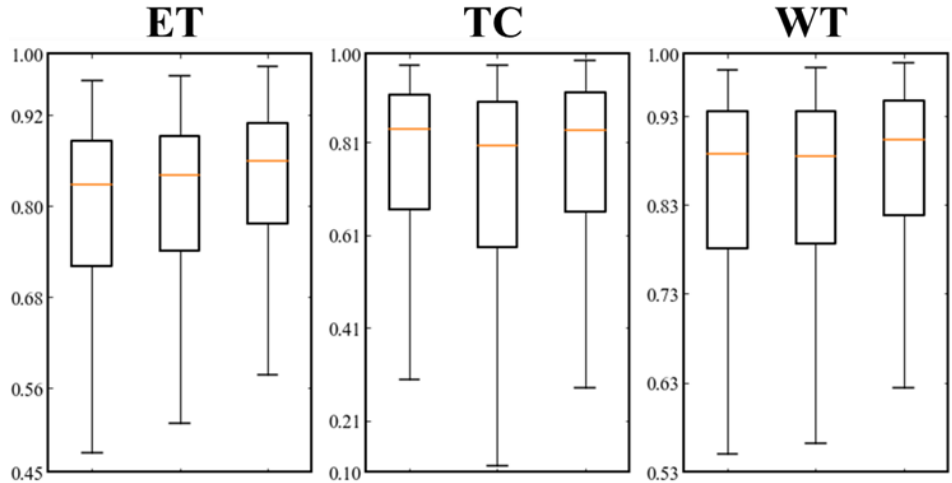

*Figure 7: The dice coefficient values boxplots given by the mp-MRI-only model, the mp-MRI + 4PCs model, and the ensemble-designed model (from the left to right)*

*Table 1. Segmentation results of the three studied methods.*
*(\* Indicates the statistically significant improvement from the mp-MRI-only result)*

|     | Inputs | mp-MRI-only | mp-MRI + 4 PCs | Ensemble Results |
| --- | --- | --- | --- | --- |
| **ET** | **Dice** | 0.777±0.153 | 0.792±0.148* | 0.817±0.136* |
|     | **Accuracy (%)** | 99.50±0.600 | 99.80±0.100* | 99.80±0.100* |
|     | **Sensitivity** | 0.822±0.207 | 0.858±0.212* | 0.929±0.146* |
|     | **Specificity** | 0.999±0.001 | 0.999±0.001 | 0.998±0.001 |
|     | Inputs | mp-MRI-only | mp-MRI + 4 PCs | Ensemble Results |
| **TC** | **Dice** | 0.742±0.237 | 0.713±0.239 | 0.757±0.212 |
|     | **Accuracy (%)** | 99.50±0.500 | 99.10±0.300 | 99.60±0.300* |
|     | **Sensitivity** | 0.896±0.178 | 0.860±0.208 | 0.940±0.141* |
|     | **Specificity** | 0.997±0.003 | 0.997±0.002 | 0.996±0.003 |
|     | Inputs | mp-MRI-only | mp-MRI + 4 PCs | Ensemble Results |
| **WT** | **Dice** | 0.823±0.165 | 0.823±0.164 | 0.854±0.143* |
|     | **Accuracy (%)** | 98.70±0.700 | 99.30±0.200* | 99.40±0.500* |
|     | **Sensitivity** | 0.888±0.187 | 0.914±0.135* | 0.931±0.128* |
|     | **Specificity** | 0.998±0.002 | 0.997±0.003 | 0.997±0.002 |

## 4. Discussion

This work proposed a novel deep ensemble learning model that incorporated radiomic feature maps (RFMs) for improving mp-MRI-based glioma segmentation accuracy. In the independent test set, the proposed model achieved an averaged dice similarity score of 0.817, 0.787, and 0.854 for ET, TC, and WT segmentation, respectively. These results are superior or very close to previously reported state-of-the-art model results in BraTS 2018-2020 challenges (Jia *et al.*, 2021; Henry *et al.*, 2021; Anand *et al.*, 2021). The incorporation of RFMs in this work is primarily driven by our hypothesis that cranial soft tissue heterogeneity information embedded in the derived image volumes may benefit mp-MRI based image segmentation. The experiment in Figure 3 revealed a strong correlation between RFM PCs and U-net deep features; thus, adding RFM PCs into U-net input variable becomes a reasonable approach. In this work, we derived 56 RFMs from each of the four MR sequences, and each patient's available image volume numbers was extended from n = 4 to n = 4+56x4 = 228. A data dimension reduction strategy applied to RFMs is thus required: in addition to the curse of dimensionality, the common deep learning working environment does not support such a big computation load. Previous radiomic feature analysis works have reported strong correlations among different feature distributions (Wu *et al.*, 2018; Barone *et al.*, 2021). One strategy is to select a portion of data with less correlations with each other; this can be facilitated by dedicated metrics such as image dissimilarity (Wang *et al.*, 2016). In this work, we adopted the PCA method since the derived PCs from orthogonal transformations can be uncorrelated with potential noise reduction effect (Jolliffe and Cadima, 2016). We successfully selected the first 4 PCs with an excellent 95% explained variance ratio. An alternative strategy is to select RFMs based on the potential correlation with U-net deep features: in our previous work of chest X-ray based RFM selection for COVID-19 diagnosis (Hu *et al.*, 2022), we calculated RFM correlations with the studied deep models' saliency maps (an attention measurement in the spatial domain). When incorporating the two RFMs with top correlation results in the input design, we successfully improved COVID-19 diagnosis accuracy. We believe that this strategy can be adapted to our current U-net based application, but it remains unclear which deep features or their derivatives would be suitable for such correlation analysis. In future research, we plan to investigate this RFM strategy through a comparative study.

The comparison of our deep ensemble model results against *mp-MRI+4 PCs* model results highlights the benefit of ensemble learning design. The four sub-models utilized different input designs that incorporated different 4 RFM PCs. Since these 4 RFM PCs were uncorrelated, the sub-model results were likely to be complementary, and thus the synthesized result could be further improved. As shown in Figures 4-6, owing to the adopted Ostu's method, the ensemble result of the ensemble segmentation result captured all key morphological details shown in the sub-model softmax probability maps. The simple averaging operation would be unable to retain the complementary image information. Additionally, the ensemble design also provides a new tool to quantify the segmentation uncertainty via examining the variation of sub-model softmax probability maps (Lakshminarayanan *et al.*, 2017). We adopted the entropy metric below to represent this data-related segmentation uncertainty:

$$U_{(i,j)} = -\sum_{t=1}^{t=4} \hat{p}_{(i,j)}^t \log_2 \left(\hat{p}_{(i,j)}^t\right) \tag{1}$$

Where $\hat{p}_{(i,j)}^t$ represents the intensity at position $(i,j)$ of the softmax probability map $t$. Figure 8 below shows an example uncertainty map (normalized to 0-100) of WT segmentation. As seen, high levels of uncertainty are observed at the edges of the WT segment, whereas the segmentation core exhibits minimal levels of uncertainty. This observation is in line with previous research on glioma segmentation uncertainty (Mehta *et al.*, 2021), which suggests that a model is considered reliable when the uncertainty is low in areas where the segmentation is accurate, while high uncertainty is expected in areas where the segmentation is inaccurate. Uncertainty quantification provides supplementary insights into a model's reliability when applied to analyze real-world cases without a ground-truth, thereby enhancing the model's confidence (Yang *et al.*, 2022). Exploring various RFM input design strategies to enhance the uncertainty estimation of deep ensemble models would be a meaningful direction for future research.

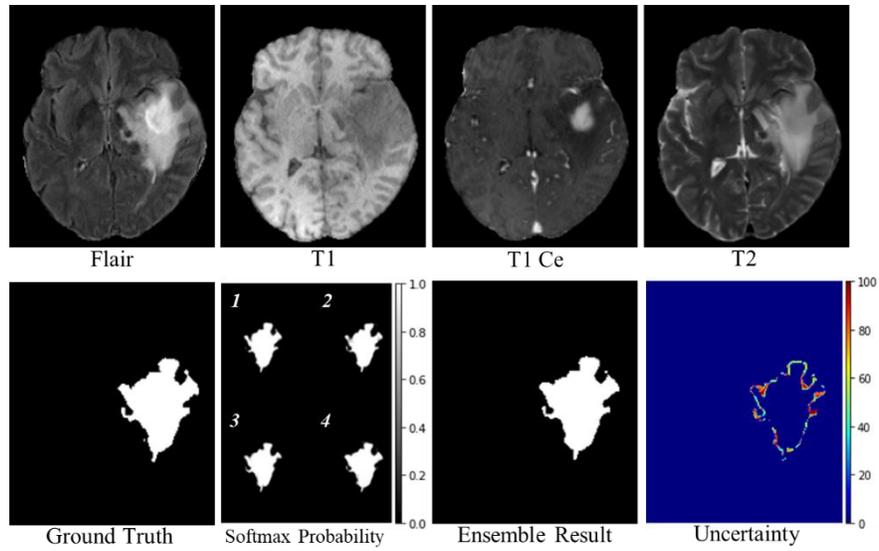

*Figure 8: An uncertainty estimation example of the WT segmentation results, the uncertainty estimation values were normalized to 0-100 for illustration purpose.*

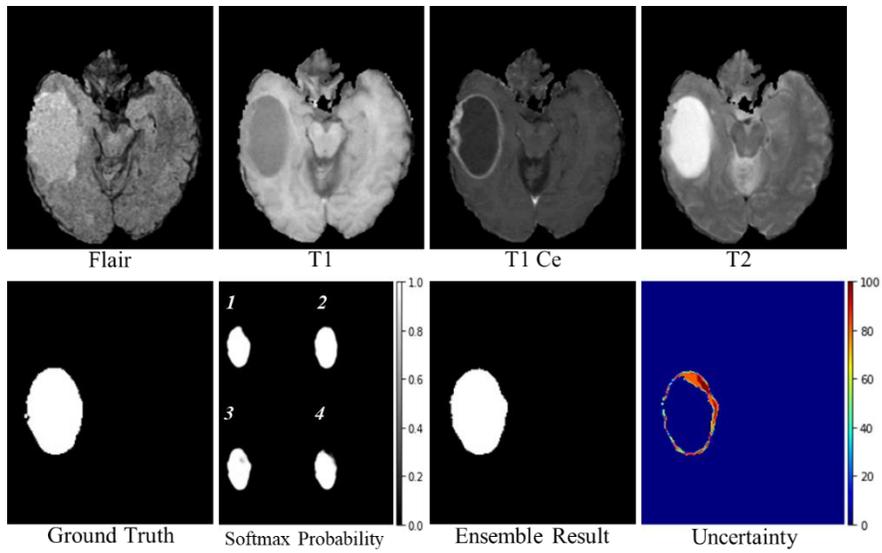

*Figure 9: An uncertainty estimation example of the TC segmentation results, the uncertainty estimation values were normalized to 0-100 for illustration purpose.*

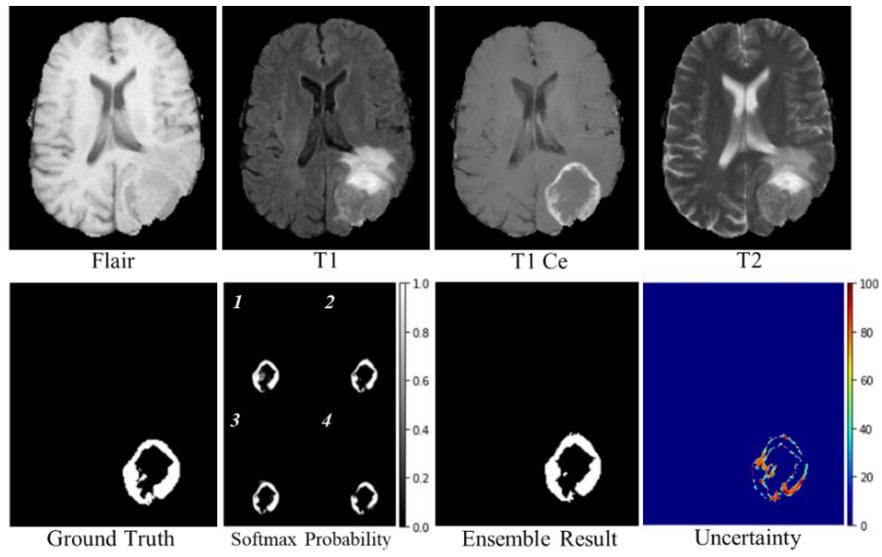

*Figure 10: An uncertainty estimation example of ET segmentation. Uncertainty values were normalized to 0-100 for illustration purpose.*

## 5. Conclusion

In this work, we have successfully developed a deep ensemble learning model that incorporates radiomics for glioma segmentation based on mp-MRI images. The model achieved improved segmentation accuracy in all three segmentation tasks we investigated, serving as a useful tool for potential clinical application. Our findings also demonstrate that the inclusion of radiomic feature maps is an effective approach that can be extended to other mp-MRI based medical image segmentation tasks.

# Reference

ABDALLA; G, ANJARI; A H M, D'ARCO; D F and BISDAS D S 2020 Glioma surveillance imaging: current strategies, shortcomings, challenges and outlook *BJR| Open* **2**

Akkus Z, Galimzianova A, Hoogi A, Rubin D L and Erickson B J 2017 Deep Learning for Brain MRI Segmentation: State of the Art and Future Directions *J Digit Imaging* **30** 449-59

Anand V K, Grampurohit S, Aurangabadkar P, Kori A, Khened M, Bhat R S and Krishnamurthi G 2021 *Brainlesion: Glioma, Multiple Sclerosis, Stroke and Traumatic Brain Injuries,* pp 310-9

Bakas; S, Reyes; M, Jakab; A, Bauer; S, MarkusRempfler;, Crimi; A and Shinohara; R T 2019 Identifying the Best Machine Learning Algorithms for Brain TumorSegmentation, Progression Assessment, and Overall SurvivalPrediction in the BRATS Challenge *arXiv:1811.02629*

Barone S, Cannella R, Comelli A, Pellegrino A, Salvaggio G, Stefano A and Vernuccio F 2021 Hybrid descriptive-inferential method for key feature selection in prostate cancer radiomics *Applied Stochastic Models in Business and Industry* **37** 961-72

Blumenthal D T, Artzi M, Liberman G, Bokstein F, Aizenstein O and Ben Bashat D 2017 Classification of High-Grade Glioma into Tumor and Nontumor Components Using Support Vector Machine *AJNR Am J Neuroradiol* **38** 908-14

Chen R, Smith-Cohn M, Cohen A L and Colman H 2017 Glioma Subclassifications and Their Clinical Significance *Neurotherapeutics* **14** 284-97

Claus E B, Walsh K M, Wiencke J K, Molinaro A M, Wiemels J L, Schildkraut J M, Bondy M L, Berger M, Jenkins R and Wrensch M 2015 Survival and low-grade glioma: the emergence of genetic information *Neurosurg Focus* **38** E6

Cybenko G, O'Leary D P and Rissanen J 1998 *The Mathematics of Information Coding, Extraction and Distribution*: Springer New York)

Deeley M A, Chen A, Datteri R, Noble J H, Cmelak A J, Donnelly E F, Malcolm A W, Moretti L, Jaboin J, Niermann K, Yang E S, Yu D S, Yei F, Koyama T, Ding G X and Dawant B M 2011 Comparison of manual and automatic segmentation methods for brain structures in the presence of space-occupying lesions: a multi-expert study *Phys Med Biol* **56** 4557-77

Deeley M A, Chen A, Datteri R D, Noble J, Cmelak A, Donnelly E, Malcolm A, Moretti L, Jaboin J, Niermann K, Yang E S, Yu D S and Dawant B M 2013 Segmentation editing improves efficiency while reducing inter-expert variation and maintaining accuracy for normal brain tissues in the presence of space-occupying lesions *Phys Med Biol* **58** 4071-97

Demirel H C and Davis J W 2018 Multiparametric magnetic resonance imaging: Overview of the technique, clinical applications in prostate biopsy and future directions *Turk J Urol* **44** 93-102

Di Ieva A, Russo C, Liu S, Jian A, Bai M Y, Qian Y and Magnussen J S 2021 Application of deep learning for automatic segmentation of brain tumors on magnetic resonance imaging: a heuristic approach in the clinical scenario *Neuroradiology* **63** 1253-62

Ghaffari M, Sowmya A and Oliver R 2020 Automated Brain Tumor Segmentation Using Multimodal Brain Scans: A Survey Based on Models Submitted to the BraTS 2012-2018 Challenges *IEEE Rev Biomed Eng* **13** 156-68

Gillies R J, Kinahan P E and Hricak H 2016 Radiomics: Images Are More than Pictures, They Are Data *Radiology* **278** 563-77

H H R M S K D I 1973 Textural features for image classification *IEEE Transactions on systems, man, and cybernetics* **SMC-3** 12

Hanif F, Muzaffar K, Perveen K, Malhi S M and Simjee Sh U 2017 Glioblastoma Multiforme: A Review of its Epidemiology and Pathogenesis through Clinical Presentation and Treatment *Asian Pac J Cancer Prev* **18** 3-9

Haralick R M 1979 Statistical and Structural Approaches to Texture *Proceedings of the IEEE* **67** 9

Henry T, Carré A, Lerousseau M, Estienne T, Robert C, Paragios N and Deutsch E 2021 *Brainlesion: Glioma, Multiple Sclerosis, Stroke and Traumatic Brain Injuries,* pp 327-39

Hesamian M H, Jia W, He X and Kennedy P 2019 Deep Learning Techniques for Medical Image Segmentation: Achievements and Challenges *J Digit Imaging* **32** 582-96

Holbrook M D, Blocker S J, Mowery Y M, Badea A, Qi Y, Xu E S, Kirsch D G, Johnson G A and Badea C T 2020 MRI-Based Deep Learning Segmentation and Radiomics of Sarcoma in Mice *Tomography* **6** 23-33

Hu Z, Yang Z, Lafata K J, Yin F-F and Wang C 2022 A radiomics-boosted deep-learning model for COVID-19 and non-COVID-19 pneumonia classification using chest x-ray images *Medical Physics* **49** 3213-22

Isensee F, Jäger P F, Full P M, Vollmuth P and Maier-Hein K H 2021 *Brainlesion: Glioma, Multiple Sclerosis, Stroke and Traumatic Brain Injuries,* pp 118-32

Işın A, Direkoğlu C and Şah M 2016 Review of MRI-based Brain Tumor Image Segmentation Using Deep Learning Methods *Procedia Computer Science* **102** 317-24

Jia H, Cai W, Huang H and Xia Y 2021 H$^2$NF-Net for Brain Tumor Segmentation Using Multimodal MR Imaging: 2nd Place Solution to BraTS Challenge 2020 Segmentation Task *Brainlesion: Glioma, Multiple Sclerosis, Stroke and Traumatic Brain Injuries* 58-68

Jiang Z, Ding C, Liu M and Tao D 2020 *Brainlesion: Glioma, Multiple Sclerosis, Stroke and Traumatic Brain Injuries,* pp 231-41

Jolliffe I T and Cadima J 2016 Principal component analysis: a review and recent developments *Philosophical transactions of the royal society A: Mathematical, Physical and Engineering Sciences* **374** 20150202

Kumar V, Gu Y, Basu S, Berglund A, Eschrich S A, Schabath M B, Forster K, Aerts H J, Dekker A, Fenstermacher D, Goldgof D B, Hall L O, Lambin P, Balagurunathan Y, Gatenby R A and Gillies R J 2012 Radiomics: the process and the challenges *Magn Reson Imaging* **30** 1234-48

Lachinov D, Vasiliev E and Turlapov V 2019 *Brainlesion: Glioma, Multiple Sclerosis, Stroke and Traumatic Brain Injuries,* pp 189-98

Lakshminarayanan B, Pritzel A and Blundell C 2017 Simple and scalable predictive uncertainty estimation using deep ensembles *Advances in neural information processing systems* **30**

Lambin P, Rios-Velazquez E, Leijenaar R, Carvalho S, van Stiphout R G, Granton P, Zegers C M, Gillies R, Boellard R, Dekker A and Aerts H J 2012 Radiomics: extracting more information from medical images using advanced feature analysis *Eur J Cancer* **48** 441-6

Lotan E, Jain R, Razavian N, Fatterpekar G M and Lui Y W 2019 State of the Art: Machine Learning Applications in Glioma Imaging *AJR Am J Roentgenol* **212** 26-37

Magadza T and Viriri S 2021 Deep Learning for Brain Tumor Segmentation: A Survey of State-of-the-Art *J Imaging* **7**

Mazzara G P, Velthuizen R P, Pearlman J L, Greenberg H M and Wagner H 2004 Brain tumor target volume determination for radiation treatment planning through automated MRI segmentation *Int J Radiat Oncol Biol Phys* **59** 300-12

Mehta R, Filos A, Baid U, Sako C, McKinley R, Rebsamen M, Dätwyler K, Meier R, Radojewski P and Murugesan G K 2021 QU-BraTS: MICCAI BraTS 2020 Challenge on Quantifying Uncertainty in Brain Tumor Segmentation--Analysis of Ranking Metrics and Benchmarking Results *arXiv preprint arXiv:2112.10074*

Otsu N 1979 A Threshold Selection Method from Gray-Level Histograms *IEEE Transactions on Systems* **9** 5

Perkuhn M, Stavrinou P, Thiele F, Shakirin G, Mohan M, Garmpis D, Kabbasch C and Borggrefe J 2018 Clinical Evaluation of a Multiparametric Deep Learning Model for Glioblastoma

Segmentation Using Heterogeneous Magnetic Resonance Imaging Data From Clinical Routine *Invest Radiol* **53** 647-54

Rizzo S, Botta F, Raimondi S, Origgi D, Fanciullo C, Morganti A G and Bellomi M 2018 Radiomics: the facts and the challenges of image analysis *Eur Radiol Exp* **2** 36

Sethi R, Allen J, Donahue B, Karajannis M, Gardner S, Wisoff J, Kunnakkat S, Mathew J, Zagzag D, Newman K and Narayana A 2011 Prospective neuraxis MRI surveillance reveals a high risk of leptomeningeal dissemination in diffuse intrinsic pontine glioma *J Neurooncol* **102** 121-7

Tiwari A, Srivastava S and Pant M 2020 Brain tumor segmentation and classification from magnetic resonance images: Review of selected methods from 2014 to 2019 *Pattern Recognition Letters* **131** 244-60

Upadhyay N and Waldman A D 2011 Conventional MRI evaluation of gliomas *Br J Radiol* **84 Spec No 2** S107-11

Wang C, Subashi E, Yin F-F and Chang Z 2016 Dynamic fractal signature dissimilarity analysis for therapeutic response assessment using dynamic contrast-enhanced MRI *Medical Physics* **43** 1335-47

Wu Y, Liu B, Wu W, Lin Y, Yang C and Wang M 2018 Grading glioma by radiomics with feature selection based on mutual information *Journal of Ambient Intelligence and Humanized Computing* **9** 1671-82

Yang Z, Lafata K, Vaios E, Hu Z, Mullikin T, Yin F-F and Wang C 2022 Quantifying U-Net Uncertainty in Multi-Parametric MRI-based Glioma Segmentation by Spherical Image Projection *arXiv preprint arXiv:2210.06512*

Yang; Z, Lafata; K, Vaios; E, Hu; Z, Mullikin; T, Yin; F-F and Wang C 2022a Quantifying U-Net Uncertainty in Multi-Parametric MRI-based Glioma Segmentation by Spherical Image Projection *arXiv:2210.06512*

Yang; Z, Lafata; K J, Chen; X, Bowsher; J, Chang; Y, Wang; C and Yin; F-F 2022b Quantification of lung function on CT images based on pulmonary radiomicfiltering *Medical Physics* **49** 9

Zhang X, Hu Y, Chen W, Huang G and Nie S 2021 3D brain glioma segmentation in MRI through integrating multiple densely connected 2D convolutional neural networks *J Zhejiang Univ Sci B* **22** 462-75